\documentclass[aps,pre,twocolumn,a4paper]{revtex4}%

\usepackage{epsfig,xcolor,soul,amsmath,amssymb,amsfonts,physics,graphicx,multirow}
\usepackage[utf8]{inputenc}
\usepackage{hyperref,csquotes}

\usepackage{xcolor}
\hypersetup{
    colorlinks=true,
    linkcolor=blue, 
    citecolor=blue, 
    urlcolor=blue, 
}

\begin{document}

\title{Interacting jammed granular systems}

\author{S\'ara \surname{L\'evay}}
\email{slevay@phy.bme.hu}
\affiliation{Department of Theoretical Physics, Budapest University of
Technology and Economics, H-1111 Budapest, Hungary}
\author{David \surname{Fischer}, Ralf \surname{Stannarius}}
\affiliation{Institute of Physics, Otto von Guericke University, D-39106 Magdeburg, Germany}
\author{Ell\'ak \surname{Somfai}, Tam\'as \surname{B\"orzs\"onyi}}
\affiliation{Institute for Solid State Physics and Optics, Wigner Research Centre for Physics, H-1121 Budapest, Hungary}
\author{Lothar \surname{Brendel}}
\affiliation{Faculty of Physics, University of Duisburg-Essen, D-47048 Duisburg, Germany}
\author{J\'anos \surname{T\"or\"ok}}
\affiliation{MTA-BME Morphodynamics Research Group, Budapest University of Technology and Economics, H-1111 Budapest, Hungary}

\begin{abstract}
More than 30 years ago Edwards and co-authors proposed a model to describe the statistics of granular packings by an ensemble of equiprobable jammed states. Experimental tests of this model remained scarce so far. We introduce a simple system to analyze statistical properties of jammed granular ensembles to test Edwards theory. Identical spheres packed in a nearly two-dimensional geometrical confinement were studied in experiments and numerical simulations. When tapped, the system evolves towards a ground state, but due to incompatible domain structures it gets trapped. Analytical calculations reproduce relatively well our simulation results, which allows us to test Edwards theory on a coupled system of two subsystems with different properties. We find that the joint system can only be described by the Edwards theory if considered as a single system due to the constraints in the stresses. The results show counterintuitive effects as in the coupled system the change in the order parameter is opposite to what is expected from the change in the compactivity.
\end{abstract} 

\date{\today; version 0.1}

\maketitle

\section{Introduction}
For a statistical description of arrangements of solid macroscopic
particles and for an analysis of the probabilities that certain states
are realized by the ensemble, Boltzmann statistics are commonly not
suitable. Granular packings are athermal, and the systems cannot
explore the configuration space by thermal fluctuations. 
In consideration of this, Edwards and Oakeshott proposed an ensemble of equiprobable jammed states to
describe granular packings~\cite{edwards1989theory}. The seemingly
contradictory concept of describing static jammed states using
equilibrium statistical physics had a mixed reception at first, but
recent advances showed the strength of it by deriving analytically the
phase space of the random packing including the packing fraction of
random close and random loose packings~\cite{baule2018edwards}.

The calculation of the partition function of the Edwards
volume~\cite{edwards1989theory,mehta1989statistical} and stress
ensemble~\cite{blumenfeld2006geometric} is difficult,
and up to now was done only in a limited number of cases. Two notable exceptions are random
packing of spheres and circles~\cite{baule2018edwards,becker2015protocol,puckett2013equilibrating,zhao2014measuring,baranau2016upper,barrat2000edwards,monasson1997entropy}, and packings in two-dimensional narrow
channels~\cite{bowles2011edwards,irastorza2013exact}. Direct
experimental and numerical verification of calculated properties are
even more scarce~\cite{puckett2013equilibrating,irastorza2013exact}.
In this paper, we consider a system where the partition function
can be expressed analytically, and our calculated expectation values of
observables agree well with the experiments and simulations.

Another important aspect which we focus on is the interaction of
jammed systems. The statistical theory of Edwards is in principle an
ideal framework for such coupled systems, but up to now there is hardly
any result regarding equilibria of jammed systems~\cite{puckett2013equilibrating,schroter2005stationary}. We will 
show that the denomination \enquote{compactivity} of the control parameter can be
misleading: In certain cases, a subsystem with higher compactivity (less compact part)
will expand rather than the connected subsystem with smaller compactivity. Nevertheless, the
interaction of the two subsystems can be described by the Edwards
ensemble but only as a whole.

\begin{figure}
\centering
\includegraphics[width=0.95\columnwidth]{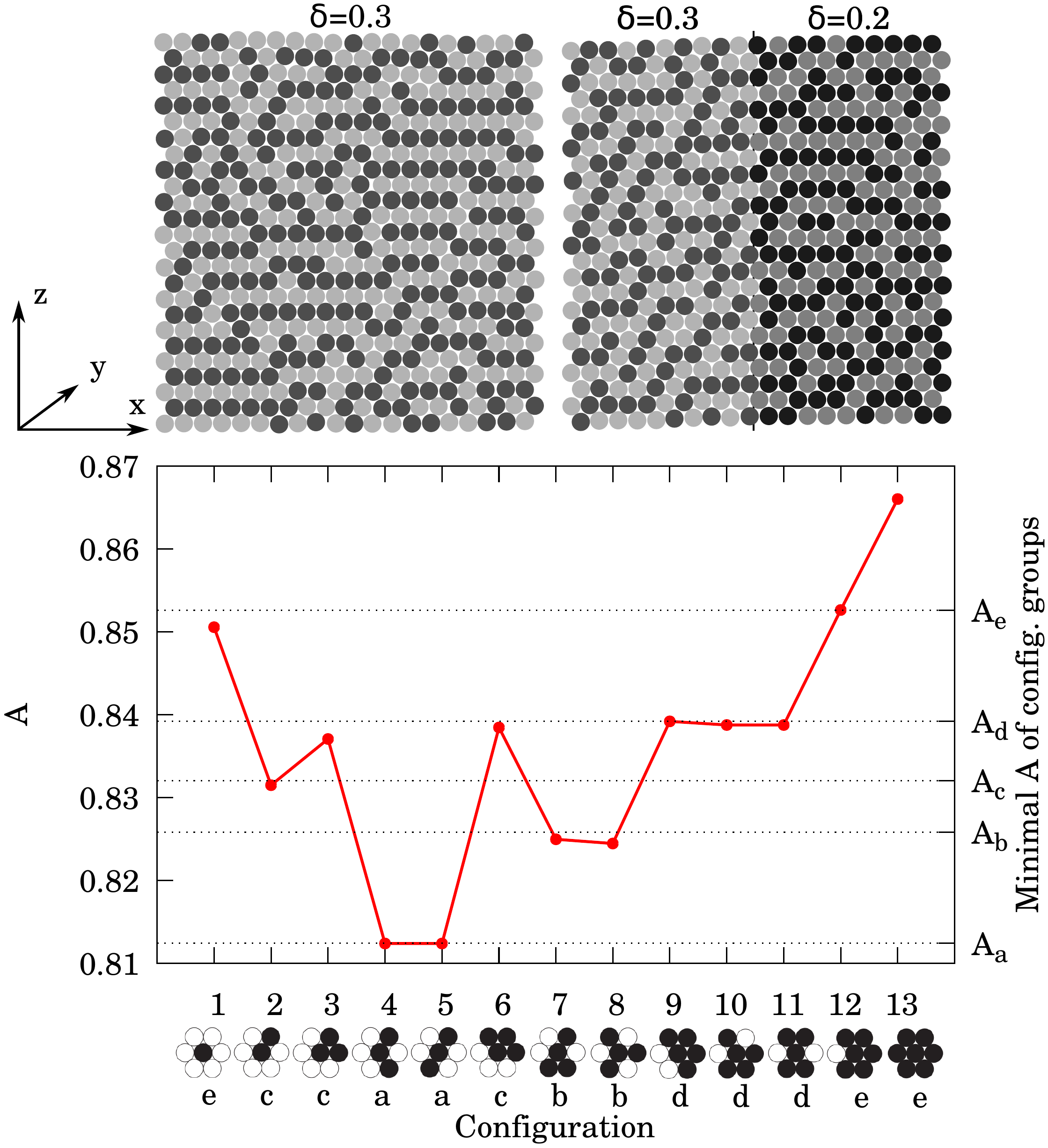}
\caption{Top: central parts of the simulated system for the normal cell (uniform width) (left) and for the coupled cell with different widths of the two sides (right). Bottom: The $13$ local configurations (see bottom line) of the packed spheres. Dark and bright circles indicate positions
at opposite cell plates. All these configuration states are degenerate; they are found twofold with the central particle positioned either at the rear or front cell wall. Most of the configurations are also found in rotated or mirrored forms. The minimal possible Voronoi area of the
central particle of the given configuration is shown by the red curve
(for $\delta{=}0.3$), calculated by kinetic Monte Carlo simulations. 
$A_i$-s on the right-hand side are the minimal Voronoi areas associated
with the five configuration groups defined in the text.}
\label{Fig:configurations}
\end{figure}

The system studied here consists of identical spheres. They are contained  in a flat cuboid with
dimensions $(L,W,H)$ in the $(x,y,z)$ directions with gravity in the
$z$ direction. Note that $H$ was much larger than the system height, allowing the system to freely shake and compactify. If the width $W{=}(1+\delta)d_p$ is only slightly larger than the
particle diameter ($d_p$), namely $0{<}\delta{<}0.45$, the ground state of the system in the $x{-}z$ plane is still a triangular lattice, although slightly distorted, with alternating
stripes of particles touching the front and back walls in the $y$ direction (see Fig.~\ref{Fig:configurations}, top).

In the following, we use dimensionless lengths, in units of the particle diameter, i.e.\ $d_p{=}1$.
Starting from a random configuration, the system begins to evolve when it is shaken periodically in the 
$z$ direction. States of the
system between the shaking periods are jammed and thus they are ideal
candidates for an Edwards ensemble. In earlier publications, it was
shown that the shaken system evolves toward the ground state but the
dynamics slows down and the configurations apparently get stuck in metastable 
states~\cite{irastorza2013exact,levay2018frustrated}. 
Snapshots of the jammed states between excitation phases are presented
in Supplemental Material (SM)~\cite{SM} (Figs. S1 and S2 and movies) for both experiments and simulations.
 Note that a substantial amount of experimental data has been collected in colloidal systems as well~\cite{han2008geometric,shokef2009stripes,shokef2011order,leoni2018attraction}.

It was shown in Ref.~\cite{levay2018frustrated} that the system can be
described by $13$ local particle configurations of a central particle
and its 6 neighbors as shown in the bottom of Fig.~\ref{Fig:configurations}. The ground state is compatible with configurations 4 and 5 only (zigzag paths or stripes of particles touching alternatively the front and back walls), so the statistical weight of these configurations ($\rho_{4,5}$) can serve as an order
parameter of the system. The different configurations have  
theoretical minimal areas (determined by sphere centers projected onto the $x{-}z$ plane), which were calculated using simulated annealing. As area, we consider the area of the Voronoi cell of the central particle of a given configuration. Areas corresponding to the tightest packings are shown in Fig.~\ref{Fig:configurations}.

The system can be treated as two dimensional, since the third dimension ($y$) is only relevant for the selection of local configurations. The global volume is determined by the $(x,z)$ positions of the particles and thus the relevant quantity is the area of the
system in the $x{-}z$ plane. The volume is considered to be 
 $(1+\delta)d_p$ times the area of a given subsystem.

\section{Results}
\subsection*{Introduction of configuration groups}
As shown in Fig.~\ref{Fig:configurations}, the minimal area of
certain subsets of configurations is (almost) the same and we can define
configuration groups to facilitate the mean field analysis as follows:
$a{=}\{4,5\}$, $b{=}\{7,8\}$, $c{=}\{2,3,6\}$, $d{=}\{9,10,11\}$,
$e{=}\{1,12,13\}$. 
(Note that configuration $13$ is practically
non-existent and $\rho_a{=}\rho_{4,5}$ is the order parameter.) The
corresponding minimal Voronoi area of the central particle in a
configuration group can be approximated by the following discrete values
using the areas $A_{1}{=}\sqrt{3}/12$ and
$A_{2}{=}\sqrt{3-4\delta^2}/12$. These are areas of triangles discussed in detail in Ref.~\cite{levay2018frustrated}: $A_1$ is the third of an equilateral triangle formed by three particles touching the same cell side, while $A_2$ corresponds to the third of an isosceles triangle with one particle located at the opposite cell side than the others. (So the Voronoi area of a perfect configuration $13$ would be $6A_1$.) The minimal area associated
with the above configuration groups can be approximated by 
$A_a{=}6A_{2}$,
$A_b{=}4.5A_{2}{+}1.5A_{1}$,
$A_c{=}3.8A_{2}{+}2.2A_{1}$,
$A_d{=}3A_{2}{+}3A_{1}$ and
$A_e{=}1.5A_{2}{+}4.5A_{1}$, shown as dotted lines in Fig.~\ref{Fig:configurations}.

\subsection*{Elementary processes during shaking}
In our study, we use the following assumption: The system is considered to form a (slightly distorted) triangular lattice in the
$x{-}z$ plane with one principal direction parallel to the $x$ axis. The spheres are touching the two particles below them~\cite{oron1998exact} (and consequently the two particles above them) and either the front or rear wall in the $y$ direction. Furthermore we will consider all volume changes up to the first order in $\delta$.

During shaking, the following processes are possible: (i) horizontal
lines gain or loose one particle, (ii) a particle changes its $y$
position (switches side), (iii) in the lowest row the particles move horizontally. Process (i) has
the highest impact on the volume of the system and happens simultaneously
in all lines generally due to global slip lines (see SM movies~\cite{SM}). Process (ii)
allows the particles to use the third dimension and optimize the
volume beyond the flat triangular lattice. This optimization is responsible
for building up the stripes in the system.  This process will create
gaps between particles which permits further compaction of the
system by allowing the next layer of particles to occupy some of the released volume. Process (iii) has no impact on the volume in first order of $\delta$ but contributes to the entropy of the system.

\subsection*{Edwards volume ensemble}
The configurational statistics were found, experimentally as well as in the simulations, to be independent of $z$ and hence of pressure (see Sect.~II in SM~\cite{SM}). Thus, we regard a row of $N$ particles as an independent sub-system described by the canonical Edwards volume ensemble and the whole system as a sample from the grand canonical ensemble. Let $L$ be the length of the container in the $x$
direction (in $d_p$ units), $M$ be the total number of particles, $K=M/N$ the number of rows, and by $\rho_i$ ($i\in\{a,b,c,d,e\}$) we denote the fraction
of different configuration groups in the system.

In order to express our partition function for a given row, we need the following quantities: volume of the system and degeneracy depending on the configuration density. The volume of the system is significantly influenced by the number of particles in a row which is changed by process (i);  process (ii) also changes the volume by optimizing configurations.

In our system the horizontal dimension is fixed. So if some free volume is available inside the system, only a part of it is eligible for compaction at the top. The extra space allows the particles to have a little horizontal gap between them in which the next layer may sink.

For the Edwards ensemble, we only need the volume change with respect to the perfect two-dimensional triangular placement of the particles. Therefore, we calculate the free space created by the above processes and we enumerate what fraction of it will be apparent at the top of the system. If $V_f$ denotes the available free space around a particle, then a simple geometric calculation yields that to first order in $\delta$, $V_g{=}V_f/(1+\sqrt{3}/2)$ will be the volume gain by the system which is visible at the top, and the volume of the Voronoi cell will be larger than the minimum by an amount of $V_f(\sqrt{3}-1/3)/(1+\sqrt{3}/2)$. So, our approach is the following: We assume that there are $K$ rows, so the perfect triangular lattice of particles would make up a volume of $V_0(K)=\frac{\sqrt{3}}{2}(1+\delta)KL$.

Naturally if $N{<}L$ then there is free space horizontally next to the particles, but that we distribute between global volume gain and the extra Voronoi volume. Thus, the free volume inside the system can be expressed as

\begin{equation}\label{eq1}
    V_{0f}(K)=\frac{\sqrt{3}}{2}(1+\delta)\left(KL-M\right),
\end{equation}
and $V_p(e)=\frac{\sqrt{3}}{2}(1+\delta)M$ denotes the volume of the $M$ particles in a perfect triangular lattice. The formula in Eq.~(\ref{eq1}) gives zero if the number of particles in a row is the same as the length of the container $N{=}L$. The advantage of working with the free volume is that both processes can be easily incorporated in the formulation.
 Process (ii) further decreases the volume of the configurations which will read as
\begin{equation}
    V_p([\rho_i])=M\sum_i\rho_iV_i.
\end{equation}
Thus the free space generated in the system is
\begin{equation}\label{Eq:Vf}
    V_{f}(K)=\frac{\sqrt{3}}{2}(1+\delta)KL-M\sum_i\rho_iV_i.
\end{equation}

Note that due to the optimization using the third dimension we may be able to put more than $L$ particles in a row, if the second term in Eq.~(\ref{Eq:Vf}) produces enough free volume.

In order to calculate the partition function we have to consider all possible configuration density distributions.
Next we have to consider the degeneracy of the systems with a given configuration density. We have two components here: First, the empty space in the first row must be distributed among the particles and then the configurations can be permuted in the system.

The first part of the degeneracy ($g_s$) is the following: The free space in the first row creates gaps between the particles which allows for their horizontal displacement. First, we calculate the degeneracy in a discretized approach assuming an elementary unit length of $\Delta \ell$. If the total gap in a line is $\ell_g$, and $k\equiv \ell_g/\Delta \ell$, then the number of ways particles can be placed in the line is
\begin{equation}
g_s(N,k,\Delta \ell) = \binom{N+k}{k}.
\end{equation}
This, of course, diverges in the limit $\Delta \ell\to 0$, but we can normalize this quantity using a well-defined system which we chose to be the ground state. Thus $\ell_{g,0}$ is the free space when we have only configurations $a$ in the system and $k'\equiv \ell_{g,0}/\Delta \ell$, so then
\begin{equation}
g_s(N,k)=\lim_{\Delta\ell\to 0} \frac{g_s(N,k,\Delta \ell)}{g_s(N,k',\Delta \ell)} =
\left(\frac{k}{k'}\right)^{N}=\left(\frac{L_f}{L_f(a)}\right)^{N},
\end{equation}
where $L_f$ is the free space horizontally. In first order in $V_f/V$, we have
\begin{equation}
    g_s(N,k) = \left(\frac{V_f}{V_f(a)}\right)^{N}.
\end{equation}

The second part of the degeneracy ($g_c$) is because the same set of configurations can be distributed in the system in many ways. Let $n_i\equiv M\rho_i$ be the number of different configurations in the system. Then, the number of different cases for positioning the different configurations in
the lattice is
\begin{equation}
g_c([n_i]) =
\frac{M!}{n_a!n_b!n_c!n_d!n_e!}.
\end{equation}

Since configurations overlap, pair correlations are extremely important; we denote by $C_{ij}$ the number of ways configuration $j$ can be placed adjacent to a given configuration from group $i$. We obtained $C_{ij}$ by generating all ($2^{19}$) possible placement of particles in a $19$ particle hexagon and counted the number of times configurations $i$ and $j$ were adjacent. This counting of adjacent configurations has the advantage that it also includes configuration degeneracy in the pair correlations, so the probability of finding a configuration $j$ next to $i$ is proportional to $C_{ij}$. A more detailed description can be found in Sec.~III of the SM~\cite{SM}.

Since all particles have 6 neighbors, we will have $3M$ neighboring particle pairs for which the probability of finding a configuration $i$ is proportional to $\rho_i$. Thus, the probability of a configuration with a given configuration density is proportional to
\begin{equation}
    \prod_{i,j}C_{ij}^{3M\rho_i\rho_j}.
\end{equation}

The grand canonical partition function up to norma\-li\-za\-tion constants is thus the following
\begin{equation}\label{Eq:zustands}
Z = \sum_N\sum_{\{\rho_i\}}g_s\,g_c
\prod_{i,j}C_{ij}^{3M\rho_i\rho_j}~
e^{V_f/X},
\end{equation}
where the positive sign in the exponential indicates that $V_f$ is the free volume the system generated on the top. Later we will also use the partition function for a system with $N$ particles in a row
\begin{equation}
    Z(N) = \sum_{\{\rho_i\}}g_s\,g_c
\prod_{i,j}C_{ij}^{3M\rho_i\rho_j}~
e^{V_f/X}.
\end{equation}

The partition sum was calculated for $N{=}6000$ particles and system length of $L{=}69$.

\subsection*{Experiments and discrete element method simulations}
We used the LIGGGHTS (LAMMPS Improved for General Granular and Granular Heat Transfer Simulations)~\cite{kloss2012models} discrete element method (DEM) simulations to
study the system. A detailed description can be found in~Ref.~\cite{levay2018frustrated} and in the Methods section.
In the cuboid cell, we simulated ${\approx}6000$ particles with periodic boundary
conditions in the $x$ direction. The width $W$ of the cell was varied
in the range $1.15~d_p~\dots~1.4~d_p$. The length of the cell was exactly
$69$ times the diameter $d_p$ of the particles. As initial conditions, we arranged
the particles into a triangular lattice with (i) random $y$ positions and
(ii) ordered $y$ positions: a striped pattern of particles touching either the
front or the rear wall. The gravitation was varied between
$1~g$ and $10~g$. In order to simulate the shaking process, particles were lifted
up and released to fall down. This resulted in different agitation energies in the range $3~mgd_p$~\dots~$100~mgd_p$ ($m$ is the particle mass). In the SM movie~\cite{SM}, snapshots of a simulation can be seen.

Reference~\cite{levay2018frustrated} and the Methods section provide a detailed description of the experimental setup as well. Images of the jammed states between shaking periods can be seen in the SM movie~\cite{SM}.

\subsection*{Calculations according to the canonical ensemble}
The first quantity we calculate is the number of particles per row:
\begin{equation}\label{Eq:avg}
\langle N \rangle=\frac{\sum_{N=1}^M N Z(N)}
{\sum_{N=1}^M Z(N)}.
\end{equation}
Surprisingly, we get $\langle N\rangle{\simeq}L$ with high accuracy
in the low compactivity regime where the experimental data can be fitted. It means that on average we should
observe a quasi two-dimensional system, which has exactly as many
particles in a row as the strictly two-dimensional system would have. In all
our simulations, we observed this law. We have performed
simulations by compressing or expanding the simulated container in
the $x$ direction and let the particles reorganize to accommodate to
the new container size, and we recovered this result. This was also the case
when we started from a perfect lattice with striped initial $y$ positions with some extra space in the $x$ direction.
The same result was observed independently of the gravity (varied in the simulations).
So from now on, we fix $N{=}L$ in all calculations, and all results presented will be done in the canonical ensemble. We have one single parameter to fit: $X$. We fit it using a single point, the order parameter. The calculation is done for different values of $X$ and we find the fitted value by interval halving which we iterate until precision $\pm0.005$ is reached on the order parameter.

\begin{figure}
\centering
\includegraphics[width=0.95\columnwidth]{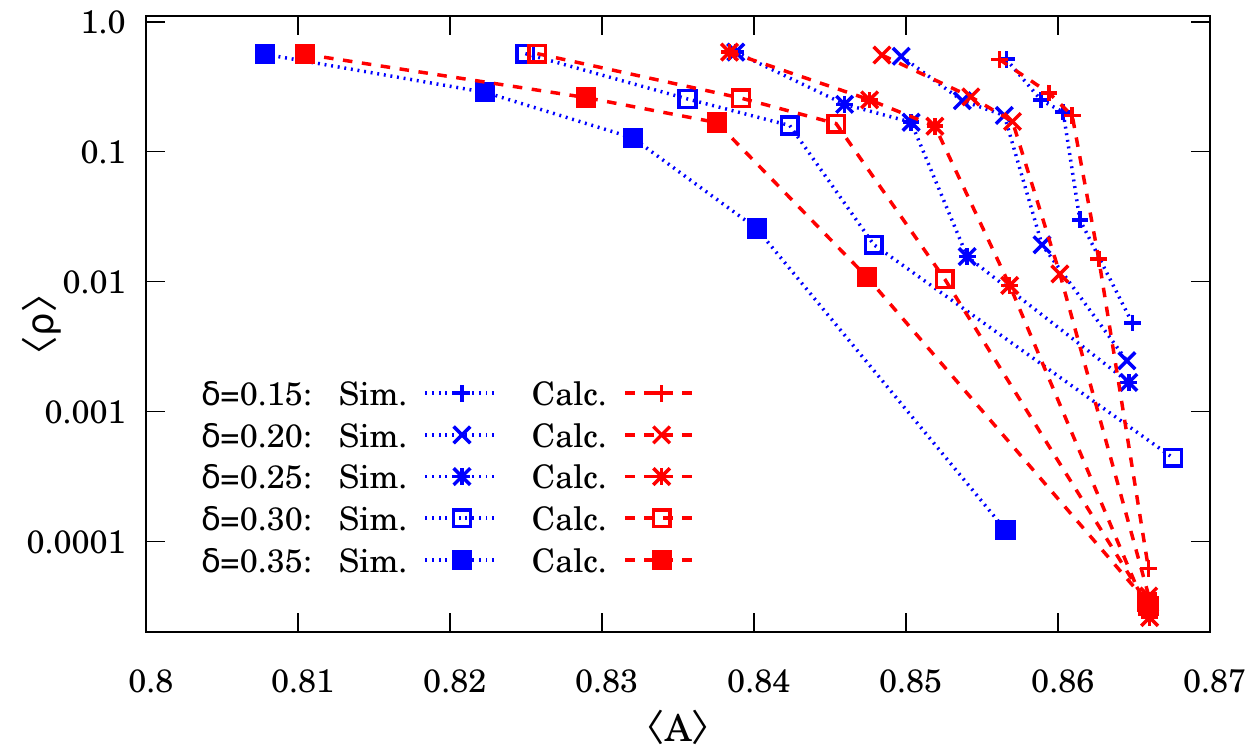}
\caption{Configuration group density vs average area for the five configuration groups for different cell thicknesses. Blue (dotted),
simulation; red (dashed), calculation for $S{=}100$ shaking periods. Fitted values of compactivity 
can be seen in Table~\ref{tab:sim_compactivity}.}
\label{Fig:Exp_Sim_Calc}
\end{figure}

\subsection*{Calculations reproducing observations and the dependence on $\delta$}
In experiments and simulations, we can determine two average quantities,
$\langle A_i\rangle$ and $\langle\rho_i\rangle$, where the former is
the average Voronoi area of configuration group $i$  including the extra
space around the central particle. The latter is simply the frequency of occurrences of the group. In Fig.~\ref{Fig:Exp_Sim_Calc}, we plot $\langle\rho_i\rangle$ versus $\langle
A_i\rangle$ for simulations for different
values of $\delta$ and for calculations which were performed with $X$ best
fitting a single point, the order parameter ($\rho_a$). One can see that the calculated
values reproduce well the observations and the dependence on $\delta$. The error increases with $\delta$ as expected since we have used free space calculations in ${\cal O}(\delta)$.

The fitted values of the compactivity $X$ in simulations (cf. Fig.~\ref{Fig:Exp_Sim_Calc}) are very different from each other. Meanwhile, the order parameter for different $\delta$ differs less than 10\% for $S{=}100$ shaking periods, as can be seen in Table~\ref{tab:sim_compactivity}. This is just a coincidence; later we will show that for longer shaking also the order parameters will be different for different $\delta$.

\begin{table}[]
    \centering
    \begin{tabular}{cccccc}
    \hline \hline
    $\delta$  & 0.15 & 0.20 & 0.25 & 0.30 & 0.35 \\
    \hline
    $X$& 0.0576& 0.0975& 0.1494 & 0.2343&0.3405 \\
    $\rho_a$ & 0.514& 0.542 & 0.583 & 0.568 & 0.561 \\ 
    \hline \hline
    \end{tabular}
    \caption{Fitted compactivities and the order parameter in case of DEM simulations with different cell widths for $S{=}100$ shaking periods.}
    \label{tab:sim_compactivity}
\end{table}

\subsection*{Incompatible domain structure can be described by Monte Carlo simulations}

We are intrigued by what sets the compactivity in our system. As the system is agitated, $X$ is decreasing, but its decrease slows down enormously around the above listed values.
Looking at snapshots of our system (see SM~\cite{SM}),  one can see that the dynamics formed domains of perfectly ordered subsystems, but these
subsystems are either incompatible with each other (different stripe
structures) or divided by seemingly stable boundary structures.

\begin{figure}
\centering
\includegraphics[width=0.95\columnwidth]{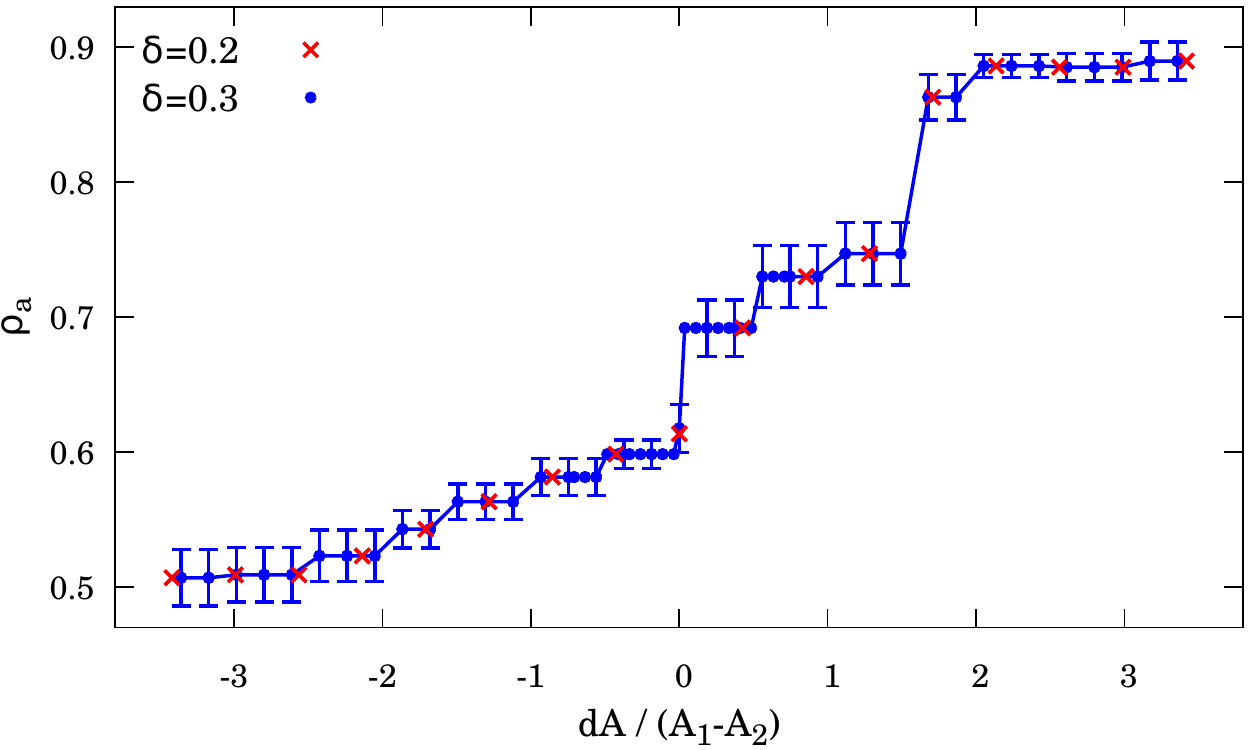}
\caption{The order parameter (the statistical weight of configurations $4$ and $5$ (configuration group $a$)) as function of the maximal allowed area
change (normalized by $A_{1}{-}A_{2}$) in the Monte Carlo model for $\delta{=}0.2$ (red) and $\delta{=}0.3$ (blue). (Each data point is an ensemble average of $10$ simulations using a given $dA$. For the blue data set, the standard deviation of different simulations is shown as well, while the red data set is shown only as a validation for different system width and thus the lesser number of data points.)}
\label{Fig:MC}
\end{figure}

It seems that the system develops a metastable domain structure
which prevents it from reaching the ground state. In order to verify this, we use the Monte Carlo model introduced in Ref.~\cite{levay2018frustrated}. (Details of the model can be found in the Methods section.) In this dynamics, we consider a system of
two state spins (spheres of front or rear position) in a triangular lattice, where the energy is
defined by the sum of the minimal Voronoi area of the resulting
configurations. The elementary step of the dynamics is a particle switch from one side of the cell to the other. Here, we ran the simulation, instead of in a
temperature controlled way, by allowing volume changes up to a maximal
limiting value. In Fig.~\ref{Fig:MC}, we show the order parameter of
the system as function of the maximum allowed volume change $dA$, normalized by the difference $A_{1}{-}A_{2}{>}0$ (which is just the difference between the area of the equilateral and isosceles triangles for a given $\delta$ discussed earlier).
$dA{<}0$ means we allow changes only if the total area is reduced by an amount greater than $|dA|$, while $dA{>}0$ means we allow slightly unfavorable changes as well: Only those changes are suppressed, which are increasing the total area by an amount greater than $dA$. One can observe a monotonically increasing stepped curve shown in Fig.~\ref{Fig:MC}. The ground state, which corresponds to $X{=}0$, is not reached at $dA{=}0$.

Surprisingly, if we flip a configuration $c$ in its average neighborhood, on average we will increase the volume of the system. Thus configurations in group $c$ are locally stable.

In order to compactify our frustrated system
further, one needs to allow unfavorable moves ($\Delta A{>}0)$. This is responsible for the slow dynamics of the system.
Furthermore, it also hinders the calculation of the compactivity of a particular system.

There is also an indication that the system is not ergodic and does not explore the whole phase space. In Ref.~\cite{irastorza2013exact}, in the 
two-dimensional version of the system the authors show that ergodicity is not observed.

\begin{figure*}
\centering
\includegraphics[width=0.7\textwidth]{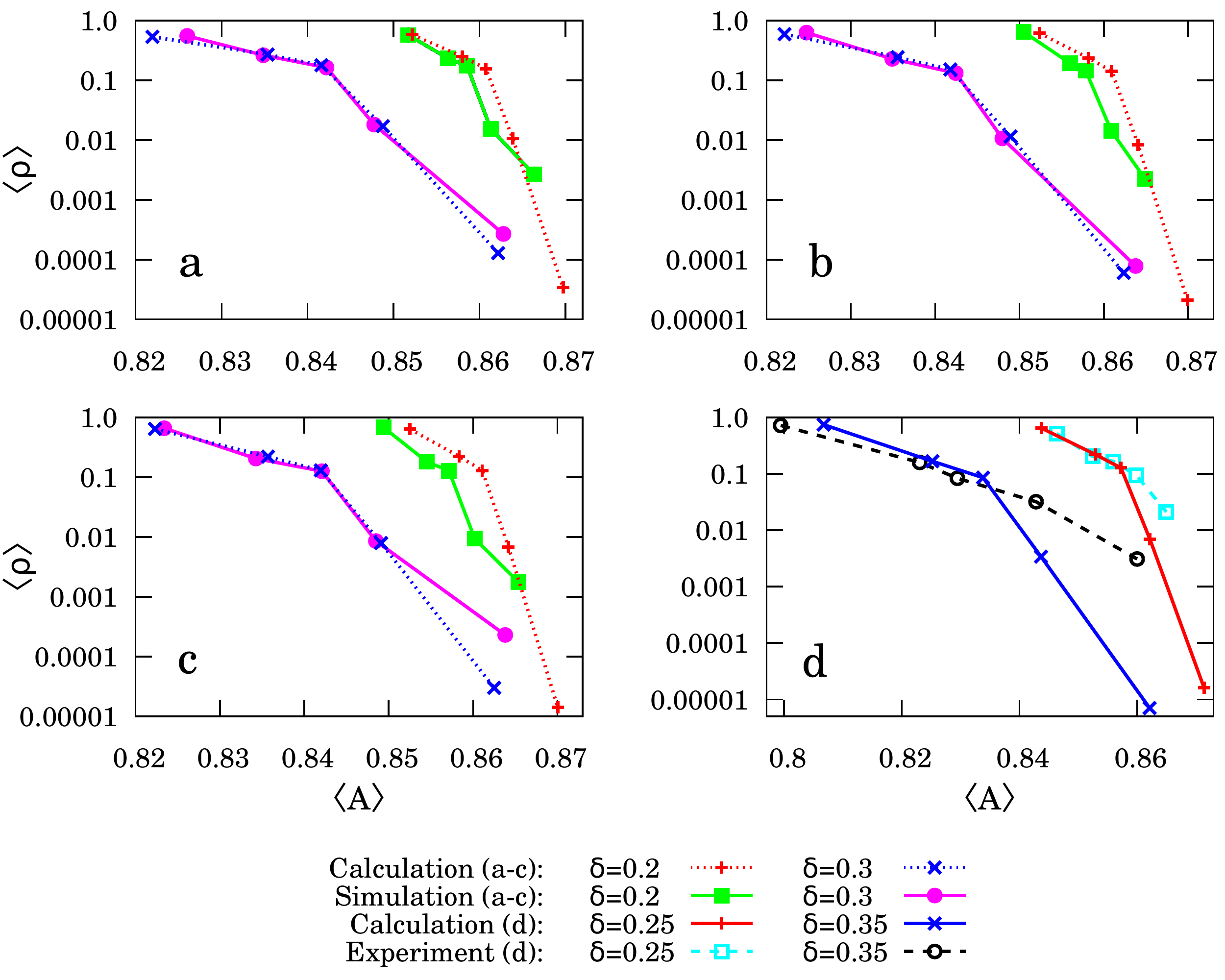}
\caption{Comparison of the configuration group density versus average area in the coupled system for the five configuration groups. [(a),(b),(c)] Comparison of simulation and calculation after $S=100, 300$, and $600$ shaking periods, respectively; (d) comparison of experiment and calculation after $S=100$ shaking periods. Values of the compactivity and order parameter can be found in Tables~\ref{tab:sim} and \ref{tab:coupled} for simulations and experiments, respectively.}
\label{Fig:side}
\end{figure*}

\subsection*{Coupling of two jammed subsystems}
Our setup allows to perform a unique experiment in which we can bring two
well-defined jammed subsystems in contact. This can
be done by changing the width ($W$) of the cell in one half of the system.
 This has been done both in experiments and in simulations (see SM movies~\cite{SM} and the top of Fig.~\ref{Fig:configurations}) with $\delta{=}0.2$ and $\delta{=}0.3$ for the different sides of the cell (in simulations, we used periodic boundary conditions in the $x$ direction).
 
 When we would like to apply the Edwards theory to this setup, we have to take into account the stress equilibrium of the two halves~\cite{baule2018edwards}. We assume that the forces between the vertical plates and the particles are negligible compared to the interparticle forces, so in principle apart from local variations we should observe hydrostatic pressure in the system which is verified in the simulations. The equilibrium between the two halves requires that we have the same height on both sides (we assume and verified that the $N{=}L$ condition still holds). Horizontally, however, one of the subsystems may gain volume on the expense of the other side. This stress equilibrium must hold for all admissible microstates~\cite{torquato2010jammed}.
 
 This feature prohibits the simple application of a common compactivity of two subsystems since the narrower side in general occupies more space than the other. One may try to shift the interface in the direction of the wider system and apply independent subsystems, but the problem remains that in this way we would consider countless microstates which violate the stress equilibrium or has negligible weight in one subsystem.
 
 The only way around this problem is to consider a joint system. We prescribe the same height on both sides for each microstate and do the same calculation as in Eq.~(\ref{Eq:zustands}), but now we use the product of two partition functions $Z_{0.2}(N/2)\times Z_{0.3}(N/2)$, with the above mentioned constraint. For the coupled system as compactivity, we chose the average of the compactivities of the standalone systems with the same number of taps. Values of the compactivity and order parameter can be found in Table~\ref{tab:sim} for DEM simulations.
 
\begin{table}[]
    \centering
    \begin{tabular}{ccp{10pt}ccp{10pt}cc}
    \hline \hline
    \multicolumn{2}{c}{System} && \multicolumn{2}{c}{Pure}
     && \multicolumn{2}{c}{Coupled}\\
     \multicolumn{2}{c}{$\delta$} && 0.2  & 0.3 && 0.2 & 0.3  \\
    \hline
    \multirow{3}{*}{$S=100$} & X && 0.0996 & 0.2355 && \multicolumn{2}{c}{0.1677} \\
     & $\rho_a$ && 0.545 & 0.565 && 0.573 & 0.554 \\
     & $\rho_a$ (calc) && 0.545 & 0.565 && 0.582 & 0.534 \\
    \hline
    \multirow{3}{*}{$S=300$} & X && 0.0984 & 0.2031 && \multicolumn{2}{c}{0.1509} \\ 
     & $\rho_a$ && 0.550 & 0.628 && 0.645 & 0.628 \\ 
     & $\rho_a$ (calc) && 0.551 & 0.628 && 0.614 & 0.592 \\
    \hline
    \multirow{3}{*}{$S=600$} & X && 0.0933 & 0.1818 && \multicolumn{2}{c}{0.1377} \\
     & $\rho_a$ && 0.573 & 0.674 && 0.679 & 0.660 \\
     & $\rho_a$ (calc) && 0.573 & 0.674 && 0.641 & 0.642 \\
     \hline \hline
    \end{tabular}
    \caption{Compactivities and the order parameter in case of DEM simulations with pure and connected cells. $S$ denotes the number of shaking periods.}
    \label{tab:sim}
\end{table}
 
 In Fig.~\ref{Fig:side}, we plot the results using the mean compactivity of the two uncoupled subsystems with same number of taps. A good match between the calculation and the numerical results can be seen [see Figs.~\ref{Fig:side}(a)-\ref{Fig:side}(c)]. Let us stress that the calculations are not fits but enumerations with the above-described joint partition function and compactivity.

 \begin{table}[]
    \centering
    \begin{tabular}{cp{10pt}ccp{10pt}ccc}
    \hline \hline
    System && \multicolumn{2}{c}{Pure}
    && \multicolumn{3}{c}{Coupled} \\
    $\delta$ && 0.25  & 0.35 & &0.25  & 0.25\footnote{In this case, during the evaluation we neglected the top $10$ layer of particles and only the lower triangular part of the remaining particles was taken into account.} &0.35  \\
    \hline
    $X$ && 0.1800 & 0.1698 & &\multicolumn{3}{c}{0.1749} \\
    $\rho_a$&& 0.505& 0.825& &0.515& 0.637& 0.722  \\
    $\rho_a$ (calc)&& 0.505& 0.825& &\multicolumn{2}{c}{0.647} & 0.744 \\
    \hline \hline
    \end{tabular}
    \caption{Compactivities and the order parameter in case of experiments with the connected cell.}
    \label{tab:coupled}
\end{table}

In the experiments the cell of the coupled system was a bit wider and had $\delta{=}0.25$ on one side and $\delta{=}0.35$ on the other. Values of the compactivity and order parameter can be found in Table~\ref{tab:coupled} for experiments.

The biggest difference between calculated and measured values were found in experiments, where we can observe much higher fraction of configuration groups $d$ and $e$ than either in the simulations or in the calculations. The reason behind this is that due to some experimental artifact we observed an extra $5{-}6\%$ particles at the back plate than at the front and thus higher frequencies of configurations with many particles at the same wall.

The other difference is that we have found only a small order parameter increase in the narrow part. In this sense, it seems that the two subsystems are not interacting as in the numerical simulation. However there is a substantial difference between the boundary conditions in the $x$ direction, which is periodic for the simulations and walls for the experiments. In the numerical simulations, the top of the system is always horizontal with some small irregularities, whereas in the experiments large slopes were also found. The reason is that the walls can support forces due to friction and thus our assumption of equal height does not hold. Since it is easier to exchange volume with the empty space above the system, this is what happens.

On the other hand, we have observed that on the narrower side the system is more ordered in a triangle (with $30^o$ angle) adjacent to the wider part. We have already reported that forces are transmitted predominantly by the lower two particles. So in this sense only configurations located in this triangular region are affected by the structure of the wider side. Indeed, in this region, we find an order parameter (denoted by $\rho_a^a$ in Table~\ref{tab:coupled}) which is compatible with the predictions of the Edwards calculation.

\section{Conclusion}
In summary, we have shown that a simple system consisting of uniform spheres in a  nearly two-dimensional cell is an excellent example to be described by the Edwards ensemble in the sense that the partition function can be formulated analytically. The observables can be calculated
exactly and the calculation matches reasonably well with simulations. 
We have shown that the system cannot reach its ground
state due to frustration in the domain structure which can only be dissolved
through unfavorable events with very small probability. Our results raise another question for a possible future study: What sets the apparent compactivity of the system?

We have also tested the applicability of the Edwards ensemble for two coupled subsystems. We found that the resulting system can only be described if the stress equilibrium is taken into account at the microstate level and the partition function of the full system is calculated. The problem of describing the coupled system as independent subsystem comes from the fact that it requires the prescribing of a previously unknown common volume distribution on both sides which is in our case an impossible task.

In summary, we have found that Edwards ensemble is capable of reproducing the observables of a jammed system but fails to help in combining subsystems when there is volume exchange not only between the subsystems but also between the subsystems and the environment.

\section{Methods}
\subsection*{Discrete Element Method simulations}
The simulations were implemented using the LIGGGHTS~\cite{kloss2012models} DEM method, consisting of a cell with sizes $(69d_p,W,{\approx}75d_p)$, where the width $W$ of the cell was varied between $1.15~d_p$ and $1.35~d_p$. Periodic boundary conditions were applied in the $x$ direction. Walls had the same mechanical and frictional properties as the grains. The cell was filled with $\approx 6000$ spherical particles with uniform diameter $d_p$. As initial filling, we applied two different methods: We arranged the particles into a triangular lattice with (i) random $y$ positions and (ii) ordered $y$ positions, a striped pattern of particles touching either the front or rear wall. (In Ref.~\cite{levay2018frustrated}, it is compared with simulations using completely random initial filling.)

In order to simulate shaking, particles were lifted up and released to fall down.
The strength of gravitation was varied between $1~g$ and $10~g$, so the agitation energy was in the range $3~mgd_p$ and $100~mgd_p$. The equilibrated configuration of particles after each shaking period is considered as a jammed state, a sample from the Edwards ensemble.
The grains are interacting when in contact via the Hertz model. The mechanical and frictional properties of particles were also varied: the coefficient of restitution between $0.25~\dots~0.75$, the coefficient of friction between $0.0~\dots~0.2$, and the Young modulus between $5\times10^6$~{Pa}~$\dots~5\times10^8$~{Pa}. Changing these parameters had no considerable effect on the results studied in this paper. A more detailed description can be found in Ref.~\cite{levay2018frustrated}, and a series of snapshots of the jammed states in the coupled cell can be seen in SM movie~\cite{SM}.

\subsection*{Experiments}
In experiments, we used vertical sandwich cells. The walls were made from glass plates and 3D printed borders. The size of the cell was $(140$~{mm}$,W,140$~{mm}$)$, and the width $W$ was varied between $1.2~d_p$ and $1.3~d_p$. For the coupled case, two transparent sheets were glued to one half of the glass plates to reduce the width of the cell. Precision glass spheres with a diameter $d_p{=}2.0{\pm}0.02$~mm were placed randomly in the cell by gravitational filling from the top. A sinusoidal signal generated by a voice coil was applied as agitation, the vertical vibration of the cell. Amplitude and frequency of the signal were varied leading to vertical accelerations between $1~g$ and $5~g$, measured by an acceleration sensor. After each shaking period, a photo from the current jammed state was taken. Uniform background illumination allowed the clear distinction of particles located at the front and rear side of the cell by their brightness. The positions of particles and resulting configurations were determined by image analysis. 
A series of images of the jammed states in the coupled cell can be seen in SM movie~\cite{SM}.

\subsection*{Monte Carlo model}
Monte Carlo simulations were performed to test the effect of mechanism (ii) of compactification (particle switches side). We were interested in the question whether the system can reach its striped ground state by particles switching sides.

To this end, we made a model where the particles were placed in a triangular lattice. The volume of the system was determined by the sum of configuration volumes of all particles as given by the minimal volume in Fig.~\ref{Fig:configurations}.

We have created a MCMC (Markov Chain Monte Carlo) algorithm using particles switching sides as elementary step and the complete volume of the system as energy. It turned out that the system compactifies more at finite temperature than at zero. To measure the volume of the necessary unfavourable elementary steps for further compaction, we have run the system instead of temperature at energy control; namely we have accepted elementary steps with volume change less than $\Delta V$. The results of the simulations are shown in Fig.~\ref{Fig:MC}.

\section*{Acknowledgments}
The authors thank
T. Trittel for support in the construction of the experimental setup. The study was funded by the Deutsche Forschungsgemeinschaft, DFG within Grants No. STA 425/38-1 and No. STA 425/46-1, by the Hungarian National Research, Development and Innovation Office (NKFIH), under Grant No. OTKA K 116036, by the BME IE-VIZ TKP2020 program, by DAAD and TEMPUS within the researcher exchange program (Grant No. 274464), and by the ÚNKP-19-3 New National Excellence Program of the Ministry for Innovation and Technology of Hungary.


\end{document}


\title{Supplemental Material for\\\textbf{\enquote{Interacting jammed granular systems}}\\by Lévay et al.}
\author{}

\maketitle

\section{Experiments and simulations}
In the following we present a few snapshots of the studied system. Systems with different width ($W{=}1+\delta$, which is the size of the cell in the $y$ dimension in diameter units) were studied as well as coupled systems, where the two sides of the cell has different width.

\begin{figure}[h]
\centering
\includegraphics[width=0.85\columnwidth]{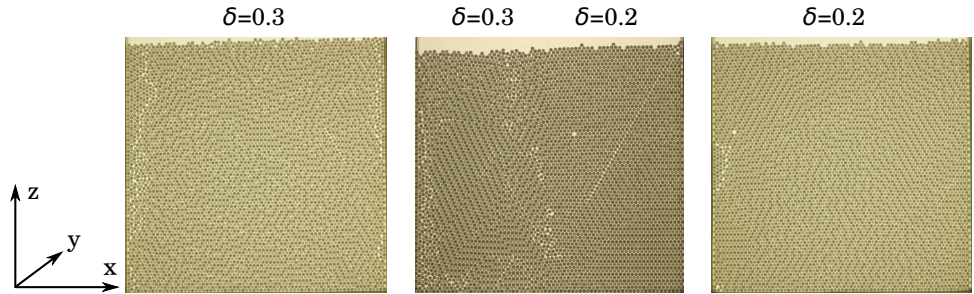}
\caption{Photos taken in experiments after $50$ shaking periods using cells with regular width with $\delta{=}0.3 \textnormal{ and } 0.2$ (on the left and the right side of the figure) and with a coupled cell (in the middle), where $\delta{=}0.3$ for the left half of the cell and $\delta{=}0.2$ for the right half. Darker particles are touching the front side of the cell while lighter particles are located at the rear side (which are physically equivalent).}
\label{Fig:exp}
\end{figure}

\begin{figure}[h]
\centering
\includegraphics[width=0.85\columnwidth]{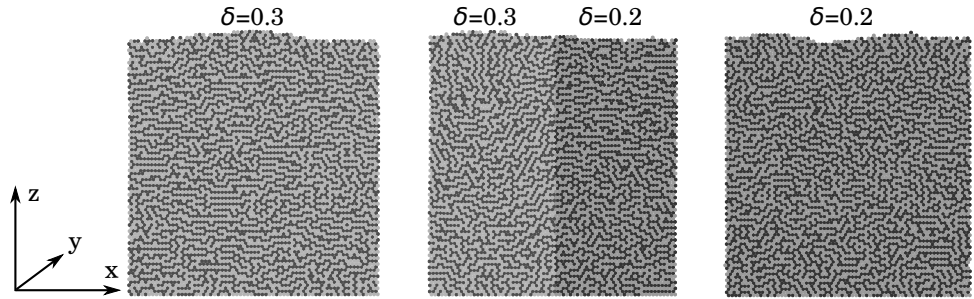}
\caption{Snapshots of DEM simulations after $50$ shaking periods using cells with regular width with $\delta{=}0.3 \textnormal{ and } 0.2$ (on the left and the right side of the figure) and with a coupled cell (in the middle), where $\delta{=}0.3$ for the left half of the cell and $\delta{=}0.2$ for the right half. Darker particles are touching the front side of the cell while lighter particles are located at the rear side. (Different shades of gray are used to facilitate the distinction between different cell widths.)}
\label{Fig:sim}
\end{figure}

In experiments photos were taken after each shaking period. A few examples can be seen in Fig.~\ref{Fig:exp}. One can also visualize snapshots from DEM simulations as can be seen in Fig.~\ref{Fig:sim}.

\section{Configurational statistics independent of $z$}
Both in experiments and simulations the configurational statistics were found independent of the $z$ coordinate, hence of pressure. Here this statement is proven by calculating the configurational statistics in parts of the system. In both cases the cell was divided to $9$ equal horizontal stripes and the configurational statistics was calculated for the parts.

\begin{figure}
\centering
\includegraphics[width=0.8\columnwidth]{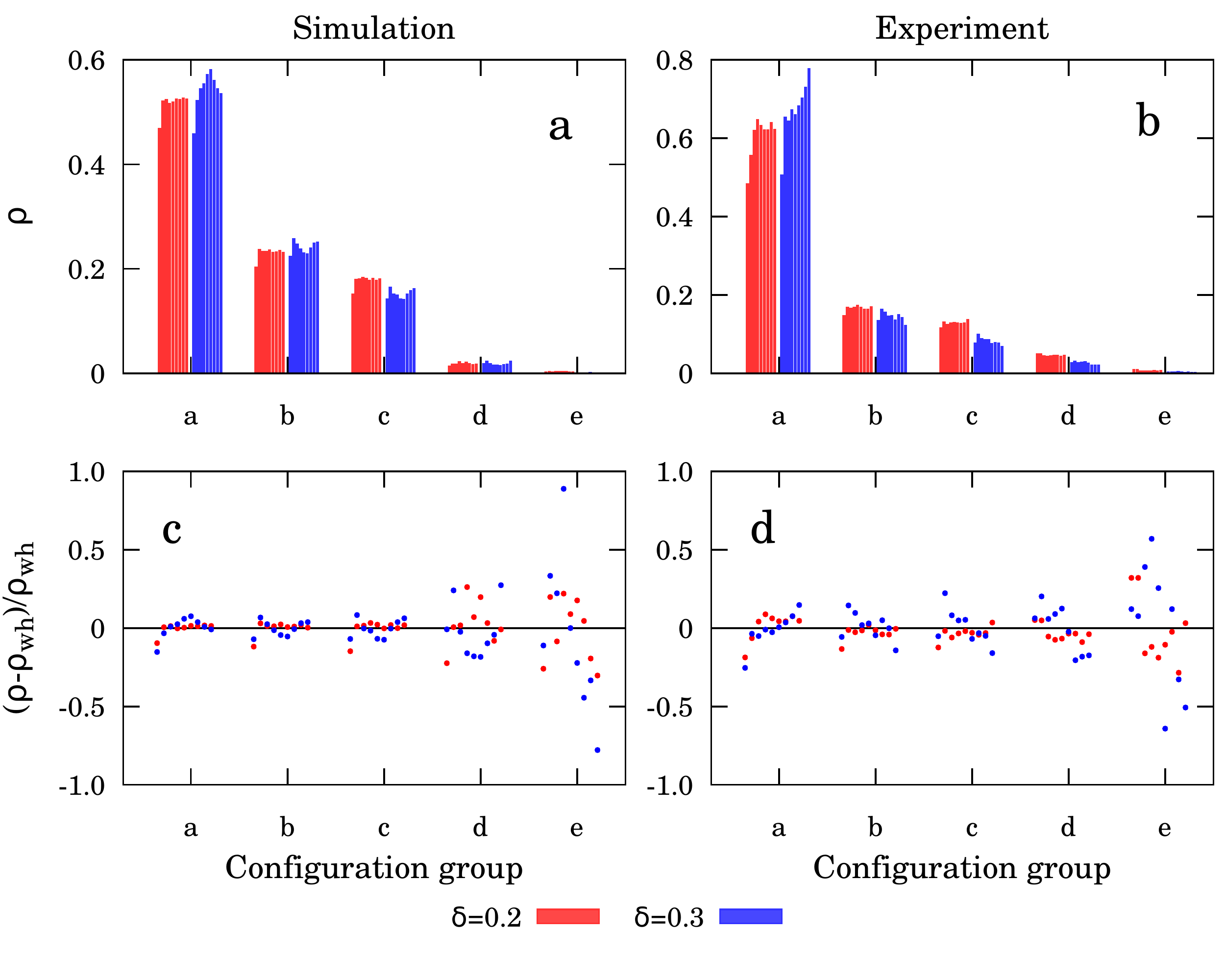}
\caption{Probability distribution of the $5$ configuration groups for $\delta{=}0.2$ (red) and $\delta{=}0.3$ (blue) for simulations (a) and experiments (b) calculated for ninths of the cell. The cell was divided to $9$ horizontal stripes (of equal size) and the statistics was calculated for these parts separately. In (c) and (d) the relative difference between the statistics of the fractions and the whole cell is presented. Narrow bars and small dots from left to right mean the statistics calculated for horizontal cell-parts from the bottom of the cell to the top. For simulations $50$ independent realizations were ensemble averaged, while for experiments we present the ensemble averaged result of $25$ independent realizations.}
\label{Fig:zIndep}
\end{figure}

The fraction of different configurations can be seen in Fig.~\ref{Fig:zIndep} (a-b) for simulations and experiments, respectively for different cell widths. In Fig.~\ref{Fig:zIndep} (c-d) the relative difference between the statistics of the fractions and the whole cell is presented. Note that the relative difference is small for configuration groups a, b, and c. These configuration groups are the most frequent, while groups d and e have very small probability to find, thus the large relative difference.

\section{Compatibility of configurations}
Since configurations overlap in the studied system, pair correlations are extremely important. We denoted by $C_{ij}$, the number of ways configuration $j$ can be placed adjacent to a given configuration from group $i$.

\begin{figure}
    \centering
    \includegraphics[width=0.25\columnwidth]{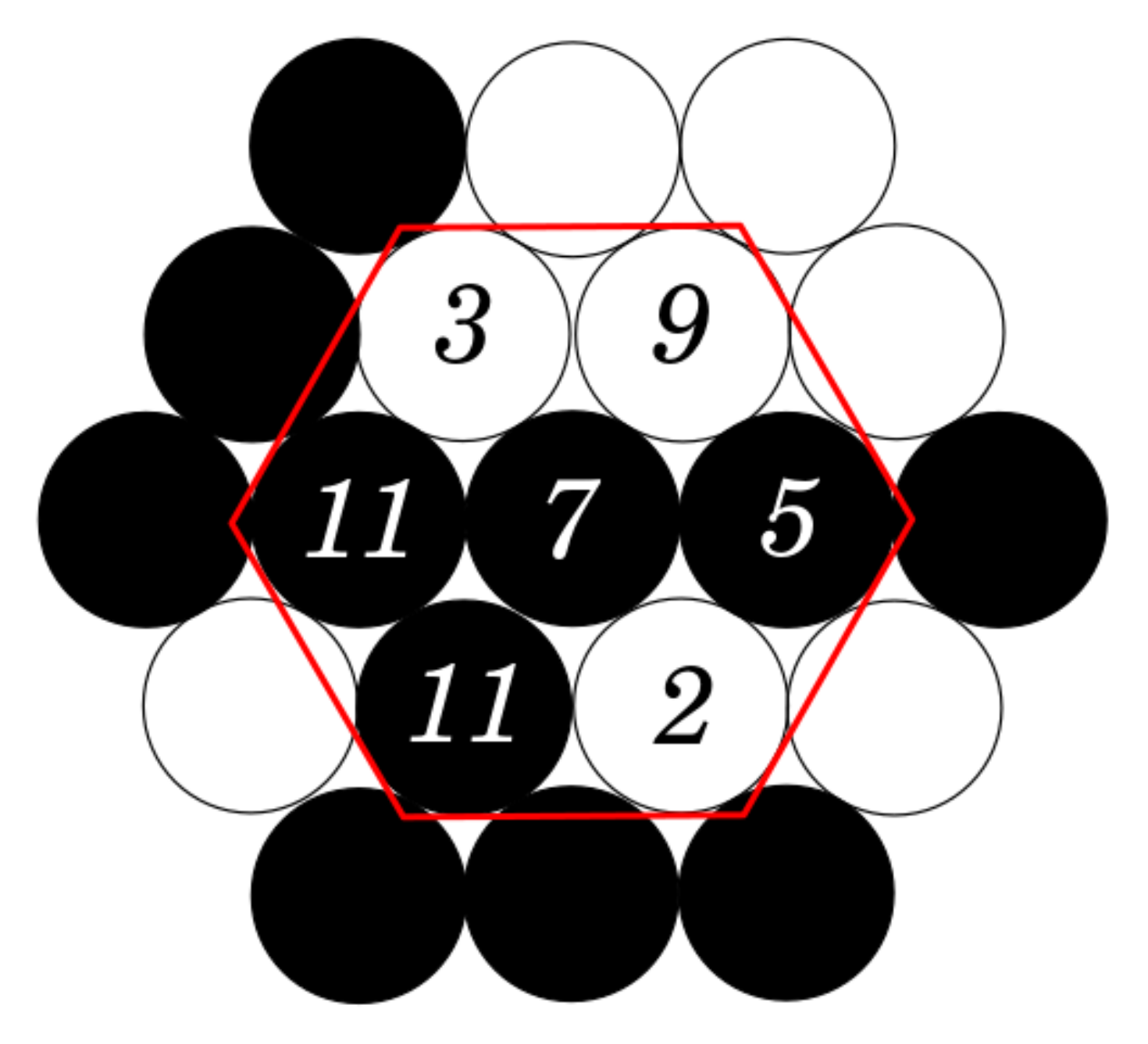}
    \caption{Part of the system, a two-shell hexagon is presented here, which contains $7$ overlapping seven-particle configurations.}
    \label{Fig:cij}
    \end{figure}

The calculation method was the following: An example of the two-shell hexagon (19 particles) can be seen in Fig.~\ref{Fig:cij}. The central configuration (red hexagon) is config. $7$ (config. group $b$), while the six neighboring (overlapping) configurations are $9 (d), 5 (a), 2 (c), 11 (d), 11 (d), 3 (c)$ (configuration numbers are written to the center of the 6-particle hexagons). The calculation of $C_{ij}$ is the following: we take every possible case for the front/back placement of the $19$ particles ($2^{19}$ possibilities) and in each case we determine the seven configurations. We then store in $C_{ij}$ the number of ways how config. $i$ can be adjacent to config. $j$ taking into account all possible placements. So the probability of finding a configuration $j$ next to $i$ is proportional to $C_{ij}$. The values obtained by calculations can be found in Table~\ref{Tab:cij}.

\begin{table}[]
    \centering
    \begin{tabular}{cp{10pt}cp{10pt}cp{10pt}cp{10pt}cp{10pt}c}
    \hline \hline
    i / j && a && b && c && d && e \\ \hline \hline
    a && 116736 && 122880 && 129024 && 61440 && 12288 \\
    b && 122880 && 153600 && 251904 && 116736 && 43008 \\
    c && 129024 && 251904 && 172032 && 245760 && 86016 \\
    d && 61440 && 116736 && 245760 && 184320 && 129024 \\
    e && 12288 && 43008 && 86016 && 129024 && 122880 \\
    \hline \hline
    \end{tabular}
    \caption{The number of ways how configurations from configuration group j can be placed adjacent to configurations from group i. Of course $C_{ij}=C_{ji}$, thus this $5\times5$ matrix is symmetric.}
    \label{Tab:cij}
\end{table}